\documentclass[12pt]{iopart}

\expandafter\let\csname equation*\endcsname\relax
\expandafter\let\csname endequation*\endcsname\relax
\usepackage{amsmath}
\usepackage{graphicx}
\usepackage{hyperref}

\begin{document}

\title{Environmental vs demographic variability in stochastic
  predator-prey models}
\author{U Dobramysl and U C T\"{a}uber}
\address{Department of Physics, Virginia Tech, Blacksburg, Virginia
  24061-0435, USA}
\ead{\mailto{ulrich.dobramysl@vt.edu}, \mailto{tauber@vt.edu}}

\begin{abstract}
  In contrast to the neutral population cycles of the deterministic
  mean-field Lotka--Volterra rate equations, including spatial
  structure and stochastic noise in models for predator-prey
  interactions yields complex spatio-temporal structures associated
  with long-lived erratic population oscillations.  Environmental
  variability in the form of quenched spatial randomness in the
  predation rates results in more localized activity patches.
  Population fluctuations in rare favorable regions in turn cause a
  remarkable increase in the asymptotic densities of both predators
  and prey~\cite{Dobramysl2008}.  Very intriguing features are found
  when variable interaction rates are affixed to individual particles
  rather than lattice sites.  Stochastic dynamics with demographic
  variability in conjunction with inheritable predation efficiencies
  generate non-trivial time evolution for the predation rate
  distributions, yet with overall essentially neutral
  optimization~\cite{Dobramysl2013}.
\end{abstract}

\paragraph{Keywords:}
\begin{itemize}
\item Population dynamics (Theory)
\item Models for evolution (Theory)
\item Mutational and evolutionary processes (Theory)
\item Stochastic particle dynamics (Theory)
\end{itemize}
\submitto{J. Stat. Mech. Theor. Exp.}

\section{Introduction}
\label{sec:introduction}

\subsection{Ecology and Population Dynamics}
\label{sec:ecol-popul-dynam}

The field of population dynamics deals with the mathematical modeling
of interacting species. It has been a very active field since about 40
years~\cite{may2001stability,Smith1974,Murray2002,Hofbauer1998} and
continues to provide exciting challenges. Ecological environments are
complicated systems with many participating agents, fluxes of energy
and resources and many inputs and outputs. There also exists a wealth
of different models for various applications. The ecological dynamics
of three, cyclically competing species of californian lizards can be
modeled using the rock-paper-scissors
model~\cite{He2010,Reichenbach2007}. In the case of highly asymmetric
interaction rates, the three species rock-paper-scissors model can be
mapped to the two-species Lotka-Volterra model~\cite{He2011}. The
cyclic competition between four and more species has also been
extensively studied~\cite{Case2010,Durney2012,Roman2012,Zia2010}.

Here, we focus on the Lotka-Volterra (LV) model, independently
introduced in 1920 by A. J. Lotka~\cite{Lotka1920}, and by V. Volterra
in 1926~\cite{Volterra1926}. The LV model consists of two species, the
predator species $A$ and the prey species $B$, obeying the following
three rules:
\begin{subequations}
  \begin{align}
    A&\stackrel{\mu}{\rightarrow}\emptyset\;, \label{eq:predator_mortality}\\
    B&\stackrel{\sigma}{\rightarrow}2B\;, \label{eq:prey_reproduction}\\
    A+B&\stackrel{\lambda}{\rightarrow}2A\;. \label{eq:predation}
  \end{align}
  \label{eq:lv_rules}
\end{subequations}
Rule \eqref{eq:predator_mortality} governs predator mortality, on its
own leading to an exponential decay of the predator population with a
characteristic rate $\mu$. Rule \eqref{eq:prey_reproduction}
represents prey reproduction and leads to an exponential increase in
the number of prey with rate $\sigma$ in the absence of any
controlling processes, such as predation or the introduction of finite
carrying capacities (i.e. restrictions on the global or local
population size). Rule \eqref{eq:predation} finally introduces
predator-prey interaction, wherein a prey particle is consumed by a
predator with a predation rate $\lambda$ and simultaneously a new
predator particle is created. Hence, the only way the predator
population can be sustained (or grow) is by consuming prey, which is
also the only way the prey population can be kept from growing
indefinitely. The simplicity of the LV model obviously leads to a
limited applicability to real ecological predator-prey systems:
\begin{itemize}
\item The prey population reproduces at a constant rate, which implies
  that growth is not limited by the availability of food resources of
  this species.
\item The mortality of single predators is uniform and does not depend
  on the abundance of prey.
\item Natural processes that might lead to prey death occur on much
  larger time scales than the predation interaction, hence they are
  negligible. This is probably justified as long as both species
  coexist. In the event of predator extinction or near-extinction this
  assumption might yield unnatural results.
\item Predator reproduction is directly coupled to predation. While it
  is reasonable to assume a connection between the reproduction rate
  of a predator species and the availability of food, a direct
  conversion of prey to predator is too simple.
\end{itemize}
This list is by no means exhaustive. A more thorough criticism of the
LV model can be found in reference \cite{Royama1971}. We nevertheless
find the LV-model to be a useful tool and a good starting point for
the study of of variability, especially due to its minimal set of
rules.

\subsection{Mean-Field Rate Equations}
\label{sec:mean-field-equations-1}

In order to construct the mean-field rate equations for the LV model
\eqref{eq:lv_rules} one assumes that the populations of both species
are well-mixed and distributed homogeneously, such that one can ignore
spatial and temporal correlations and fluctuations. Since the predator
population decreases exponentially with rate $\mu$, the change of the
predator population has to include the term $-\mu a(t)$, where $a(t)$
denotes the time-dependent spatially averaged density of species $A$. 
Similarly, the prey population density $b(t)$ increases exponentially 
with a rate $\sigma$, hence its first derivative must include the term 
$\sigma b(t)$. The predation interaction depends on the availability 
of both predators and prey, hence the interaction term has to depend 
on both densities and the predation rate $\lambda$. The interaction is
conservative in the sense that one prey is converted into exactly one 
predator, thus the mean-field (mass action) factorization of the 
interaction term  $\lambda a(t) b(t)$ enters positively and negatively
into the predator and prey density change, respectively. Putting 
everything together yields the LV mean-field rate equations:
\begin{subequations}
  \begin{align}
    \frac{da(t)}{dt}&=\lambda a(t) b(t) - \mu a(t)\;, 
    \label{eq:lv_mean_field_predator}\\
    \frac{db(t)}{dt}&=\sigma b(t) - \lambda a(t) b(t)\;. 
    \label{eq:lv_mean_field_prey}
  \end{align}
  \label{eq:lv_mean_field}
\end{subequations}
These equations may be derived in a more formal manner via the master
equation of the LV model~\eqref{eq:lv_master_equation}
(for the detailed procedure see section~\ref{sec:stoch-simul} below
and reference~\cite{Tauber2012}).

\begin{figure}[t]
  \centering
  \includegraphics[width=0.9\columnwidth]{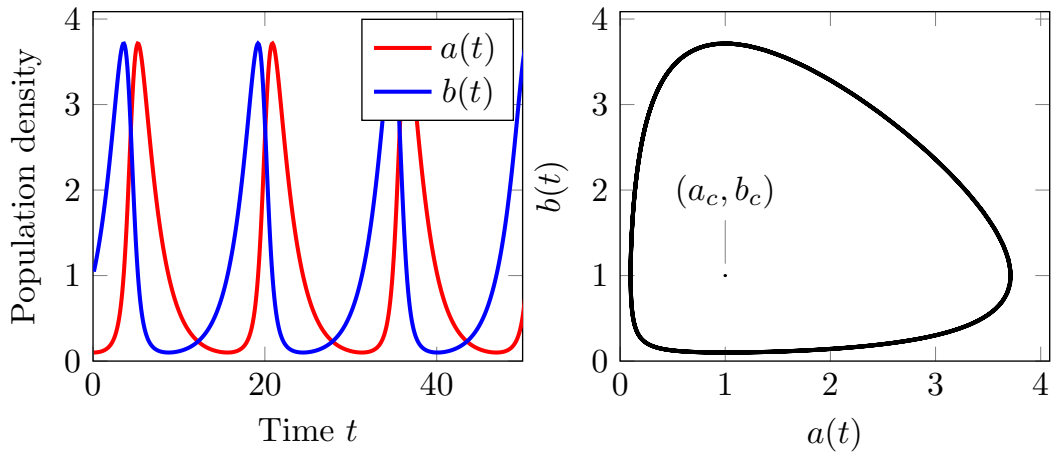}
  \caption[LV mean-field oscillations and phase space
  trajectory.]{LV mean-field oscillations and phase space
    trajectory. The LV mean-field rate equations
    \eqref{eq:lv_mean_field} give rise to nonlinear oscillations
    around the coexistence fixed point
    $(\sigma/\lambda,\mu/\lambda)$. The left panel shows the predator
    and prey population densities $a(t)$ and $b(t)$ as a function of
    time $t$ for the reaction probabilities $\sigma/\tau=0.5$,
    $\mu/\tau=0.5$ and $\lambda/\tau=0.5$ ($\tau$ indicates unit time)
    and initial densities of $a(0)=0.1$ and $b(0)=1$. The oscillations
    are clearly visible. The right panel displays the phase space
    trajectory as a closed cycle around the coexistence fixed point.}
  \label{fig:lv_mean_field_osc}
\end{figure}
By setting the left-hand side of
equations~\eqref{eq:lv_mean_field_predator}
and~\eqref{eq:lv_mean_field_prey} to zero, one immediately finds the
fixed points of this system with stationary densities $(a_{fp},
b_{fp})$:
\begin{enumerate}
\item The trivial fixed point where both population densities are zero
  $(0,0)$.
\item The predator extinction fixed point, with the prey population
  tending to infinity $(0,\infty)$.
\item The species coexistence fixed point where both predator and prey
  densities are finite $(a_c=\sigma/\lambda,b_c=\mu/\lambda)$.
\end{enumerate}
The trivial and the predator extinction fixed points are both linearly
unstable with respect to small perturbations in the densities $a$ and
$b$ (with $\lambda=0$ the state $(0,\infty)$ becomes stable). The
coexistence fixed point is marginally stable: linear stability 
analysis yields purely imaginary eigenvalues, hence this fixed point 
gives rise to marginal population cycles. This is the origin of the 
characteristic LV oscillations, displayed in figure 
\ref{fig:lv_mean_field_osc}; in the limit of small amplitudes, the
linear oscillation frequency is $\omega=\sqrt{\mu\sigma}$. These 
stable phase space orbits are associated with a first integral of 
motion of the LV mean-field rate equations. By dividing
equations~\eqref{eq:lv_mean_field_predator}
and~\eqref{eq:lv_mean_field_prey} and separating the variables we get
\[\Bigl(\frac{\sigma}{a}-\lambda\Bigr)da
=\Bigl(-\frac{\mu}{b}+\lambda\Bigr)db\;.\]
Integrating both sides yields the constant expression
\begin{equation}
  \label{eq:lv_first_integral}
  K(t)=\sigma\ln a(t)+\mu\ln b(t)-\lambda[a(t)+b(t)]=K(0)\;,
\end{equation}
which is also the Lyapunov function of the LV system. A rigorous 
stability analysis of the LV mean-field equations can be found in 
references~\cite{Mobilia2006a,Washenberger2007}.

\subsection{Stochasticity and Simulations}
\label{sec:stoch-simul}

\begin{figure}[t]
  \centering
  \includegraphics[width=0.9\columnwidth]{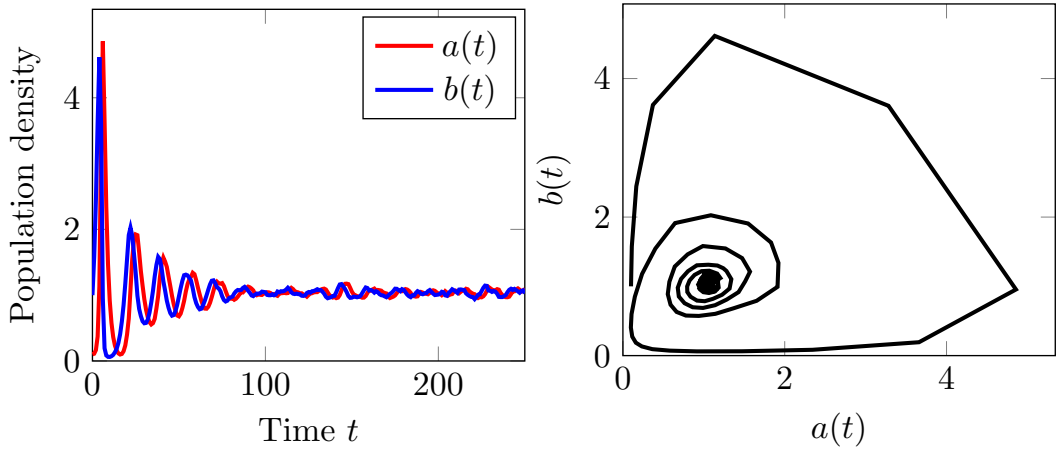}
  \caption[Population densities from a single LV Monte
  Carlo simulation run.]{Population densities from a single Monte
    Carlo simulation run. The left panel shows the predator and prey
    densities $a(t)$ and $b(t)$ as functions of time $t$ for 
    $\sigma=0.5$, $\mu=0.5$, and $\lambda=0.5$, with initial densities
    $a(0)=0.1$ and $b(0)=1$. One can clearly see that the LV 
    oscillations are now damped due to the stochastic nature of the 
    simulation; the densities approach the coexistence fixed point. 
    After reaching the steady state the densities perform erratic 
    oscillations around these values driven by population number 
    fluctuations. The right panel shows that the phase space 
    trajectory of the simulation run is a spiral beginning at the 
    initial densities of $(0.1,1)$ and approaching the fixed point 
    $(1+\epsilon_a,1+\epsilon_b)$. The deviations $\epsilon_a$ and 
    $\epsilon_b$ stem from a renormalization of the mean-field 
    steady-state densities due to fluctuations.}
  \label{fig:lv_spatial_stochastic_osc}
\end{figure}
The derivation of the mean-field equations discussed in the last
section assumes that the system is well-mixed and deterministic. This
assumption is in general not valid for ecological systems, which are
stochastic in nature. Hence, we numerically solve the underlying
master equation using Monte Carlo simulations. Via a rescaling with an
appropriate unit of time we obtain the single-particle probabilities
from the rates $\mu$, $\sigma$ and $\lambda$. The master equation of
the LV system with parallel and independent updates reads:
\begin{equation}
  \label{eq:lv_master_equation}
  \begin{split}
    \frac{dP(A,B,t)}{dt}=&\mu (A+1) P(A+1,B,t)+\sigma (B-1) P(A,B-1,t)\\
    &+\lambda (A-1)(B+1)P(A-1,B+1,t)-(\mu A+\sigma B+\lambda A B)P(A,B,t)\;,
  \end{split}
\end{equation}
where $P(A,B,t)$ denotes the probability of the system being in a
state with $A$ predator and $B$ prey particles at time $t$. By solving
equation \eqref{eq:lv_master_equation} for $dP/dt=0$ one finds that
this system has exactly one steady-state solution, namely
$P(A=0,B=0,t\rightarrow\infty)=1$ and
$P(A>0,B>0,t\rightarrow\infty)=0$~\cite{Haken1983}. Hence, the
previously unstable trivial extinction fixed point becomes a stable
absorbing state. Due to the discrete nature of the stochastic system,
fluctuations in the number of particles can drive the population into
extinction if the number of particles becomes small. Moreover, this
result implies that any system with a finite number of particles
\emph{always} reaches the extinction state, but the extinction time
scale can become quite large already for reasonably sized
systems~\cite{Rulands2013}. This feature is absent in the mean-field
rate equations since the population densities can get arbitrarily small
without the population going extinct. The marginally stable species
coexistence fixed point in the mean-field model becomes metastable in
a stochastic system. While fluctuations will eventually drive the
system to extinction, the coexistence state is long-lived. In Monte
Carlo simulations, the population densities approach the coexistence
densities via damped oscillations starting from the initial
conditions. Figure \ref{fig:lv_spatial_stochastic_osc} shows the
species densities over time and the resulting spiral in phase space
for a representative simulation run. Population fluctuations lead to
small oscillations around the steady-state densities after the system
reached the stationary state. Internal white noise stemming from the
demographic stochasticity excites the resonant frequency of the system
and results in these small oscillations. This leads to a drastic delay
in the ultimate extinction of the system~\cite{McKane2005}.

By introducing the mean particle densities as
\[a(t)=\sum_{A,B=0}^\infty AP(A,B,t)\qquad\text{and}\qquad
b(t)=\sum_{A,B=0}^\infty BP(A,B,t)\;,\] we can derive the mean-field
rate equations~\eqref{eq:lv_mean_field}. Taking the time derivative of
the mean predator density and inserting the master
equation~\eqref{eq:lv_master_equation} yields
\begin{equation*}
  \begin{split}
    \frac{da(t)}{dt}=&\sum_{A,B=0}^\infty A\frac{d}{dt}P(A,B,t)\\
    =&\sum_{A,B=0}^\infty\bigl[\mu A(A+1)P(A+1,B,t)+\sigma A(B-1)P(A,B-1,t)\\
    &+\lambda A(A-1)(B+1)P(A,B,t)-(\mu A+\sigma B+\lambda AB)AP(A,B,t)\bigr]\\
    =&\sum_{A,B=0}^\infty\bigl[-\mu A+\lambda AB\bigr]P(A,B,t)\;,
  \end{split}
\end{equation*}
where we shifted the summations over $A$ and $B$ in the last step. In
order to arrive at the mean-field rate equation, we need to make the
approximation that the probability to be in a state described by the
particle numbers $A$ and $B$ factorizes into the independent
probabilities of having $A$ predators and $B$ prey, $P(A,B,t)\approx
P(A,t)P(B,t)$. This leads to
\[\frac{da(t)}{dt}\approx-\mu a(t)+\lambda a(t)b(t)\;,\]
which is identical to the mean-field rate equation for the predator
population~\eqref{eq:lv_mean_field_predator}. An analogous derivation
yields the rate equation describing the prey
population~\eqref{eq:lv_mean_field_prey}.

\subsection{Spatial Structure}
\label{sec:spatial-structure}

\begin{figure}[tp]
  \centering
  \includegraphics[width=0.9\columnwidth]{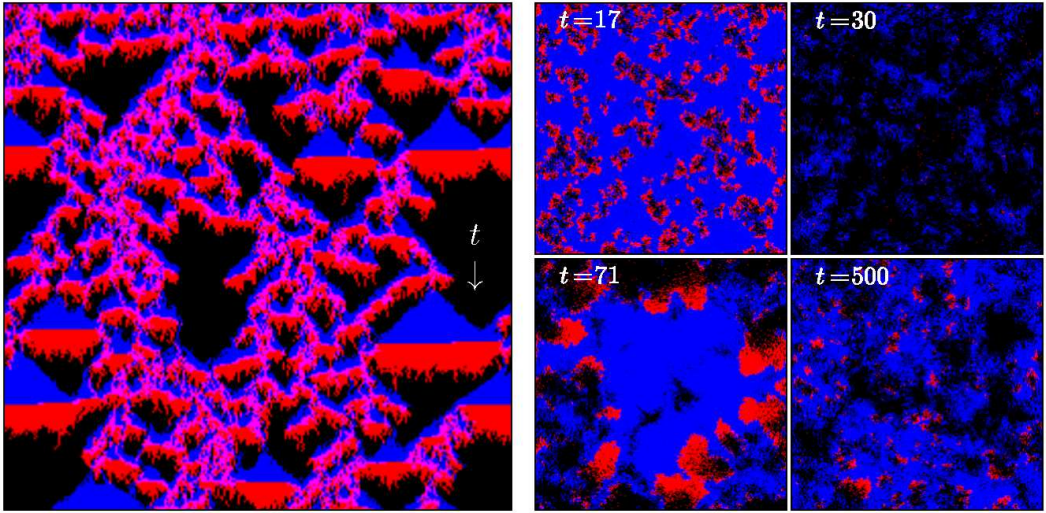}
  \caption[Snapshots from a one-dimensional and a two-dimensional
  spatial LV Monte Carlo simulation run.]{Snapshots from a
    one-dimensional and a two-dimensional spatial LV Monte
    Carlo simulation run. The left panel shows a one-dimensional
    simulation with $L=250$, $\mu=0.5$, $\sigma=0.5$ and
    $\lambda=0.3$, and initial densities $a(0)=1=b(0)$. The vertical
    direction shows time evolution while the horizontal is the spatial
    direction. The colors blue and red indicate the presence of prey
    and predator particles respectively, while a black pixel indicates
    an empty site. At $t=0$ the system is still well-mixed and
    clusters of prey particles form and grow over time. Predators
    invade prey clusters and thereby often remove them completely. 
    The right panel displays several snapshots from a two-dimensional 
    simulation with $L=250$, $\mu=0.9$, $\sigma=0.1$, and $\lambda=1$. 
    In the initial configuration, particles were randomly distributed 
    with densities $a(0)=0.01$ and $b(0)=1$. The predator-prey 
    community survives an initial predator invasion ($t=17$), which 
    leads to a subsequent prey proliferation due to predator scarcity 
    ($t=30$). Predator fronts start to invade a large prey cluster 
    ($t=71$). After the initial transient oscillations, the system 
    reaches the coexistence quasi-steady state characterized by 
    smaller prey clusters and predator invasion fronts ($t=500$).}
  \label{fig:lv_spatial_fronts_snapshot}
\end{figure}
Ecological systems exhibit spatial structure. Members of species move
through the environment foraging or evading predators. This leads to
spatial correlations in the abundance of species, and emerging spatial
patterns such as spirals or
wavefronts~\cite{Murray2002,Reichenbach2007,Dunbar1983,He2011b}. None
of these features are captured by a mean-field model or by
zero-dimensional stochastic models. It is however sometimes possible
to use a stochastic PDE model to describe spatial
patterns~\cite{Reichenbach2008}. In Monte Carlo simulations in an
ecological context, one generally uses a simple hyper-cubic lattice of
edge length $L$ and dimensionality $d$ to introduce spatial
structure. It is assumed that a simple diffusion process is adequate
to describe the movement of species through space. Hence, particles
hop from one lattice site to a randomly chosen neighboring site,
performing random walks. A remarkable feature of the LV model is that
the results in the coexistence phase are qualitatively independent of
the details of the simulation
method~\cite{Mobilia2006a,Washenberger2007,Mobilia2006}. Yet it should
be noted that the introduction of global or local population number
restrictions induces a predator extinction threshold, separating the
two-species coexistence phase from a state with proliferating prey
filling the entire system. For the predator population this phase
represents an inactive, absorbing state. Throughout this article, the
number of particles is essentially unrestricted (up to a safety limit
that is never reached in our simulations), hence this phase does not
exist in our models.

Figure \ref{fig:lv_spatial_fronts_snapshot} shows representative
snapshots from a one-dimensional and a two-dimensional LV Monte Carlo
simulation. The one-dimensional simulation (left-hand panel)
progresses by forming prey clusters that are subsequently invaded by
predators, which leads to intriguing spatio-temporal patterns. In the
two-dimensional simulation, patches of prey particles form and become
invaded by predators. Initially several large clusters span the
system, which yields the observed synchronized oscillations in the
densities. As the simulations progress the clusters become smaller and
more numerous, hence the invasion cluster growth cycles de-synchronize
throughout the lattice and the quasi-steady state is reached. The
right-hand side of figure \ref{fig:lv_spatial_fronts_snapshot} shows
several snapshots from a representative simulation run.

The remainder of this paper is structured in the following way: In the
next section, we set the stage by briefly summarizing our findings on
environmental variability in the interaction rates. In
section~\ref{sec:demogr-vari}, we discuss a non-spatial stochastic
model for demographic variability and introduce our approximate
mean-field description. We also compare the non-spatial model with
results from two-dimensional simulations. In
section~\ref{sec:spat-vs.-demogr}, we investigate the full spatial
system which includes both environmental and demographic
variability. We finally conclude with section~\ref{sec:conclusions}.

\section{Environmental Variability}
\label{sec:envir-vari}
\begin{figure}
  \centering
  \includegraphics[width=0.9\columnwidth]{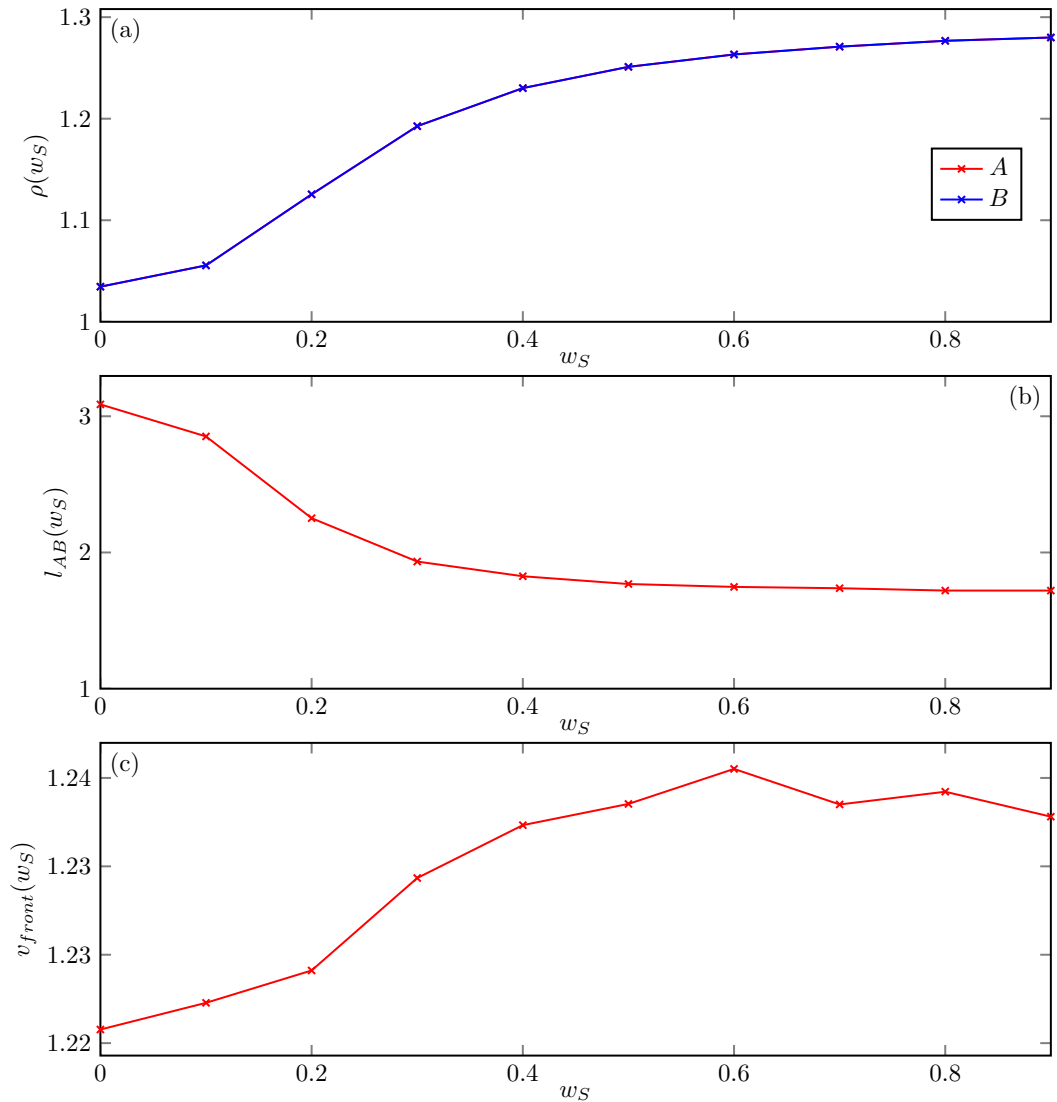}  
  \caption{Effects of environmental variability. (a) The predator and
    prey species densities are enhanced by up to $24\%$, as a function
    of the variability $w_S$. (b) The typical distance between
    predator and prey particles decreases with increasing
    variability. (c) The speed of traveling wavefronts is enhanced as
    well. Data in this figure are taken from
    reference~\cite{Dobramysl2008}.}
  \label{fig:envir-vari}
\end{figure}

Ecological systems are in general not homogeneous. The availability of
energy and resources can vary significantly over various length
scales. In most ecological models, the effects of environmental
variability are assumed to only enter via a trivial renormalization of
the coarse-grained reaction rates. However, if the variation takes
place on a similar length scale as the interactions, its effects are
not adequately captured by such a simplified description. Hence, in
our spatial ecosystem model, we choose to describe spatial variability
by making the reaction rates spatially distributed quenched random
variables, subject to a Gaussian probability distribution, truncated
to the interval $[0,1]$, with mean $\frac{1}{2}$ and standard
deviation $w_S$. This standard deviation is a measure of the amount of
environmental variability and is thus a model parameter. More details 
about the implementation are given in reference~\cite{Dobramysl2008}.

Our previous work on this model confirmed the crucial role of
environmental variability on the particle abundance. We found that the
densities of \emph{both} predator and prey species increase by up to
$24\%$ as a function of the variation strength; see
figure~\ref{fig:envir-vari}(a). Other quantities are affected as well:
e.g., the relaxation time to reach the quasi-steady state and the
correlation lengths for either species as well as their typical 
separation distance become reduced with increasing disorder variance. 
This led us to conclude that the overall population increase was 
caused by more narrowly localized activity patches in which prey 
proliferate and predators feed off the out-diffusing prey; c.f. 
figure~\ref{fig:envir-vari}(b). Variability in the other rates $\mu$ 
and $\sigma$ did not lead to significant changes in the system; see 
reference~\cite{Dobramysl2008} for a detailed discussion. In our
investigation, we also measured the speed of traveling activity fronts
$v_{front}$ (i.e. the fronts formed by a predator species invading a
prey population). We found a small but significant enhancement of this
quantity; see figure~\ref{fig:envir-vari}(c).

Environmental variability has also been subsequently
investigated in the context of cyclic models, particularly the three
species rock-paper-scissors and May-Leonard models. The effects of
spatial variations in the reaction rates on both of these models were
surprisingly small, which indicates that cyclic three-species models
seem to be robust against the introduction of environmental
variability~\cite{He2010,He2011b}. The macroscopic properties of these
systems are hardly modified by stochastic fluctuations in general.

\section{Demographic Variability}
\label{sec:demogr-vari}

Variability can also be considered in the context of variation between
individual members of each species. Due to differences in genetic 
heritage and learned strategies, the effectiveness with respect to 
reproduction, death, predation, etc. can vary between individuals of 
the same species. Hence, we may view the efficiency at certain 
processes as properties or traits of individual agents when modeling 
these systems. We focus again on the non-linear predation process and 
render the predation rate of a particular interaction between a 
predator and a prey particle a function of their respective predation
efficacies.

The investigation of individual or demographic variability directly
leads into a discussion of population-level evolution and optimization
of traits. It is reasonable to assume that offspring inherit certain
abilities from their parents. These abilities can be derived from the
genetic make-up that is inherited from the parent generation, or they 
could also be strategies for food gathering or hunting patterns, 
learned through imitation from their immediate social surroundings. A 
combination of these determines a particular individual's efficiency 
at a given process, whence the more discrete nature of the genetic 
make-up and the presence or absence of certain strategies is smeared 
out. This coarse-grained interpretation of process efficiencies 
finally allows us to assume that the efficacy value of a given 
offspring particle will be situated in the vicinity of its parent's. 
The severity of genetic mutations as well as the accuracy of strategy
imitation between generations then determines the amount of 
inheritance variability of the coarse-grained efficiencies. Applied to
the previously introduced LV system, this scheme allows for specific
optimization of predation efficacies at the level of species 
populations, as discussed in subsection~\ref{sec:popul-distr}.

Optimization and evolution in predator-prey systems has been studied
previously in experimental and theoretical contexts, by means of
different models: Kishida {\em et al.} investigated reciprocal 
phenotype plasticity in salamanders and its tadpole prey. The gape of 
the salamander species adapted as a function of the body size of the
tadpoles~\cite{Kishida2006}. Yoshida {\em et al.} studied prey
evolution in an experimental model using planktonic rotifers, and
modeled this system using a system of nonlinear differential
equations~\cite{Yoshida2003}. Fort and Inchausti employed an 
agent-based model that included a niche axis to study the emergence 
of biodiversity~\cite{Fort2012}. Rogers {\em et al.} designed a niche 
model and applied a master equation expansion, showing that 
demographic noise leads to the spontaneous formation of
species~\cite{Rogers2012}. Traulsen {\em et al.} investigated
evolutionary dynamics in unstructured populations using a stochastic
differential equations approach~\cite{Traulsen2012}. Weitz {\em et 
al.} studied the co-evolution of bacteria and bacteriaphage via
mean-field and stochastic models~\cite{Weitz2005}. While our
investigation was partially motivated by these previous studies, our
focus is different, namely on the influence of demographic 
variability on systems that exhibit the potential of evolutionary 
optimization.

We define our model in the following subsection~\ref{sec:model-rules}.
In subsection~\ref{sec:mean-field-equations}, we derive the associated
mean-field equations and discuss their steady-state solutions. 
Finally, subsection~\ref{sec:popul-distr} deals with the results of
non-spatial as well as two-dimensional stochastic simulations and a
comparison with the mean-field approximation.

\subsection{Model Rules}
\label{sec:model-rules}

We use the LV model as a basis for our study of individual 
variability, as explained already in section~\ref{sec:introduction}.
In our model, particles of either species have an intrinsic property 
that describes their efficacy during predation reactions. More 
specifically, each particle carries a predation efficiency value 
$\eta$ between zero and one. During a predation interaction between 
predator and prey particles with respective efficiency values $\eta_A$
and $\eta_B$, we choose to determine the actual reaction rate from the
arithmetic mean of these efficiencies:
\begin{equation}
  \label{eq:lambda_efficiencies}
  \lambda=\frac{1}{2}(\eta_A+\eta_B)\,.
\end{equation}
Consequently, a predator particle with a high predation efficacy has a
higher chance of consuming a prey and reproducing; it can be 
considered a good ``hunter''. A prey particle with a low efficiency
value is generally less likely to be consumed and can hence be labeled
good at ``evading''. Note that this efficiency value $\eta$ differs 
from the fitness value that is derived from a certain genotype, which 
is often defined as the average number of offspring. Our net predation 
efficacy is a mesoscopic continuous stochastic variable that describes 
the combined effects of genetic makeup and strategy learning on the 
hunting or evasion capabilities of each individual.

\begin{figure}[t]
  \centering
  \includegraphics[width=0.7\columnwidth]{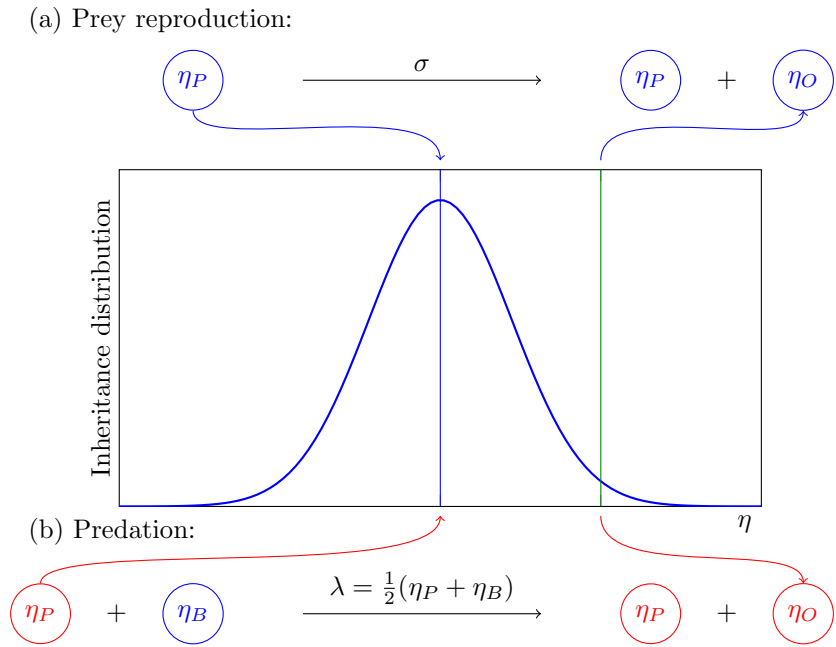}
  \caption[Inheritance model rules.]{Inheritance model rules. (a)
    During prey reproduction, a parent particle spawns an offspring
    prey particle with a rate $\sigma$. The parent particle's
    predation efficiency value $\eta_P$ is used as the mean of a
    Gaussian distribution, truncated to the interval between zero and
    one. The offspring's efficiency value $\eta_O$ is then drawn from
    this inheritance distribution. (b) During predation, a predator
    particle consumes a prey particle with a rate $\lambda$, a
    function of the participating particles' predation efficiency
    values. A new offspring predator particle is created and its
    efficiency value $\eta_O$ is determined via the same mechanism as
    is used in the case of prey reproduction.}
  \label{fig:inheritance_rule}
\end{figure}
We include inheritance and thus evolutionary dynamics in the predation
efficiencies. We argued earlier that the predation efficiency value of
an offspring particle is likely to be near the parent's value. Since
the prime goal of this work is to investigate the influence of
variability, we need to be able to control the average efficiency
deviation between generations (i.e. the mutation probability). This
suggests the use of a symmetric probability distribution that exhibits
a maximum around the parent particle's efficiency value with a
well-defined second moment.

Figure~\ref{fig:inheritance_rule} shows how the predation efficiency
is determined during the reproduction processes of prey and predator
particles. The parent particle's efficiency value $\eta_P$ is used as
the mean value of a Gaussian distribution, truncated to the interval
$[0,1]$. The offspring's efficacy value $\eta_O$ is then drawn from
this inheritance distribution. The standard deviation of the Gaussian
distribution (before truncation) $w_P$ serves as the measure of
variability in this scheme. It can also be viewed as the average
severity of mutations from one generation to the next.

This variable inheritance of efficacies now allows for evolution of
the hunting and evasion capabilities of the predator and prey
populations. Selection processes due to the predation reaction will
optimize the steady-state population distributions of {\em both} 
species. There is however a notable asymmetry between the optimization
mechanisms for the predator and prey populations, since the species 
interaction directly affects only the predator reproduction, whereas 
the prey population optimization happens through an indirect selection
bias. We shall discuss this point further in 
section~\ref{sec:popul-distr}.

\subsection{Doi-Peliti Formalism and Mean-Field Equations}
\label{sec:mean-field-equations}

In this subsection we systematically derive the mean-field equations
for the LV model with inheritance of efficiencies. We start by writing
down the model's master equation, which describes the time evolution
of the probabilities of the system's microscopic states. We then
switch to an equivalent Fock space formulation using particle creation
and annihilation operators, which allows us to rewrite the master
equation as an ``imaginary-time Schr\"{o}dinger equation''. This
yields a Liouville (or pseudo-Hamiltonian) time evolution operator. We
write down the coherent-state 'action' in terms of the ladder operator
eigenvalues and finally arrive at the mean-field equations for the
predator and prey particle numbers. Their steady-state solutions can
then be found numerically. Finally, we derive the exact solution in
the case of a uniform inheritance distribution.

\subsubsection{Master Equation.}
\label{sec:master-equation}
\newcommand{\cu}[1]{\{#1\}}

To construct the master equation of our LV-system with demographic
variability and inheritance of a continuous efficiency variable, we
need to find an equivalent system with a discrete set of states. To
this end, we discretize the interval of possible predation efficiency
values $0\le\eta\le1$ into $N$ bins, with the bin midpoint values
$\eta_i=(i+1/2)/N,\,i=0...N-1$. We then consider a predator or prey
particle with an efficacy value in the range
$\eta_i-1/2\le\eta<\eta_i+1/2$ to belong to the predator or prey
subspecies $i$. The probability for the system to be in a state with a
collection of $\cu{n_0,...,n_{N-1}}\equiv\cu{n}$ particles of
subspecies of type A and $\cu{m_0,...,m_{N-1}}\equiv\cu{m}$ particles 
of subspecies of type B at time $t$ is given by $P(\cu{n},\cu{m},t)$. 
In the following, the notation $\cu{n_i+1}$ indicates that there are
$\cu{n_0,n_1,...,n_i+1,...,n_{N-1}}$ particles in the collection.

The probability that a particle with predation efficiency $\eta_1$
produces offspring with efficiency $\eta_2$ will be assigned using a
reproduction probability function $f(\eta_1,\eta_2)$. We do not make
any assumptions about the shape of this probability distribution other
than that it be symmetric under exchange of its arguments, and that it
be properly normalized with $\int_0^1d\eta_1f(\eta_1,\eta_2)=1$. We 
use the discretized form $f_{ij}=f(\eta_i,\eta_j)$. The quantity
$\lambda_{ij}=(\eta_i+\eta_j)/2$ finally provides the interaction rate
of particles $i$ and $j$.

Gathering the inflow and outflow terms of all reactions, we arrive at
the master equation of the LV system with demographic variability and
evolutionary dynamics:
\begin{equation}
  \label{eq:mastereq}
  \begin{split}
    &\frac{\partial P(\cu{n},\cu{m},t)}{\partial t}=
    \mu\sum_{i}[(n_i+1)P(\cu{n_i+1},\cu{m},t)-n_iP(\cu{n},\cu{m},t)]\\
    &\;+\sigma\sum_{i}\Bigl[\sum_{k}(m_i-\delta_{ik})f_{ik}
    P(\cu{n},\cu{m_k-1},t)-m_iP(\cu{n},\cu{m},t)\Bigr]\\
    &\;+\sum_{i}\sum_{j}\lambda_{ij}\Bigl[\sum_{k}(n_i-\delta_{ik})
    (m_j+1)P(\cu{n_k-1},\cu{m_j+1},t)-n_im_jP(\cu{n},\cu{m},t)\Bigr]\;.
  \end{split}
\end{equation}

As initial probability distribution $P(\cu{n},\cu{m},t_0)$, we choose
independent Poisson distributions for both particle subspecies,
\begin{equation}
  \label{eq:poisson-distrib}
  P(\cu{n},\cu{m},t_0)=\Bigl(\prod_i\frac{\overline{n}_0^{n_i}}{n_i!}
  e^{-\overline{n}_0}\Bigr)\Bigl(\prod_j\frac{\overline{m}_0^{m_j}}
  {m_j!}e^{-\overline{m}_0}\Bigr)\;,
\end{equation}
where the mean initial predator and prey species densities are 
denoted as $\overline{n}_0$ and $\overline{m}_0$.

\subsubsection{Equivalent Fock Space Formulation and the Time Evolution
  Operator.}
\label{sec:hamiltonian}
\newcommand{\bra}[1]{\left|#1\right>}
\newcommand{\ket}[1]{\left<#1\right|}

Because transitions between states of this system are uniquely 
identified by integer changes in the occupation numbers of the
subspecies, we can introduce a general state $\bra{\phi(t)}$ through 
the linear combination
\begin{equation}
  \label{eq:genstate}
  \bra{\phi(t)}=\sum_{\cu{n}}P(\cu{n},\cu{m},t)\bra{\cu{n},\cu{m}}\;,
\end{equation}
of all possible basis states
$\bra{\cu{n},\cu{m}}=\prod_{i}\bra{n_i}\bra{m_i}$ for our system,
weighted by the configurational probability of each state. By
differentiating with respect to time, inserting the master equation
(\ref{eq:mastereq}) and shifting the summation over states, we obtain
\begin{equation*}
  \begin{split}
    \frac{\partial \bra{\phi(t)}}{\partial t}=& \sum_{\cu{n}}
    \sum_{\cu{m}}P(\cu{n},\cu{m},t)\Bigl(\mu\sum_in_i\bigl[
    \bra{\cu{n_i-1},\cu{m}}-\bra{\cu{n},\cu{m}}\bigr]\\
    &+\sigma\sum_im_i\Bigl[\sum_kf_{ik}\bra{\cu{n},\cu{m_k-1}}
    -\bra{\cu{n},\cu{m}}\Bigr]\\
    &+\sum_i\sum_j\lambda_{ij}n_im_j\Bigl[\sum_kf_{ik}\bra{\cu{n_k+1},
    \cu{m_j-1}}-\bra{\cu{n},\cu{m}}\Bigr]\Bigl)\;.
  \end{split}
\end{equation*}

Next we introduce raising and lowering operators in complete analogy 
to a quantum-mechanical harmonic oscillator or bosonic Fock states. We
need two sets of operators, $a_i$, $a^{\dagger}_i$ and $b_i$,
$b^{\dagger}_i$ for species A and B, respectively. The operators act
on the states in the following manner:
\begin{equation}
  \label{eq:raislower}
  \begin{split}
    a_i^\dagger\bra{\cu{n},\cu{m}}=&\bra{\cu{n_i+1},\cu{m}}\;,\\
    a_i\bra{\cu{n},\cu{m}}=&n_i\bra{\cu{n_i-1},\cu{m}}\;,\\
    b_i^\dagger\bra{\cu{n},\cu{m}}=&\bra{\cu{n},\cu{m_i+1}}\;,\\
    b_i\bra{\cu{n},\cu{m}}=&m_i\bra{\cu{n},\cu{m_i-1}}\;,\\
    \bigl[a_i,a_j^\dagger\bigr]=&\delta_{ij}=\bigl[b_i,b_j^\dagger\bigr]\;,
  \end{split}
\end{equation}
guaranteeing that the occupation number operators $a_i^\dagger a_i$ and
$b_i^\dagger b_i$ have integer eigenvalues. This procedure finally 
yields the time evolution or Liouville operator of our system
\begin{equation}
  \label{eq:lv_hamiltonian}
  H=\sum_i\biggl[\mu(a_i^\dagger-1)a_i+\sigma\Bigl(1-\sum_kf_{ik}
  b_k^\dagger\Bigr)b_i^\dagger b_i+\sum_j\lambda_{ij}\Bigl(b_j^\dagger
  -\sum_kf_{ik}a_k^\dagger\Bigr)a_i^\dagger a_ib_j\biggr]\,.
\end{equation}
Note that one may obtain this result from the reaction Liouville
operator of the standard LV system~\cite{Tauber2012} by replacing
$1-b_i^\dagger\to1-\sum_kf_{ik}b_k^\dagger$ in the prey reproduction
term, as well as 
$b_j^\dagger-a_j^\dagger\to b_j^\dagger-\sum_kf_{jk}a_k^\dagger$ and
$\lambda\to\sum_j\lambda_{ij}$ (with the appropriate change in the
indices) in the predator reproduction term, as one would expect.

\subsubsection{Coherent-State 'Action' and Mean-Field Equations.}
\label{sec:meanfield}

To calculate observable averages $\left<O(t)\right>=
\sum_{\cu{n}}\sum_{\cu{m}}O(\cu{n},\cu{m})P(\cu{n},\cu{m},t)$ one needs 
to introduce a projection state $\ket{P}=\ket{0}\prod_ie^{a_i}e^{b_i}$ 
with $\left<{P}\middle|0\right>=1$ and 
$\ket{P}a_j^\dagger=\ket{P}=\ket{P}b_j^\dagger$ due to
$\bigl[e^{a_i},a_j^\dagger\bigr]=e^{a_i}\delta_{ij}$~\cite{Tauber2012}. 
We can then write observable averages as
\begin{equation}
  \label{eq:observables}
  \left<O(t)\right>=\ket{P}O(\cu{n},\cu{m})\bra{\phi(t)}\,.
\end{equation}
Due to probability conservation
$1=\left<P\middle|\phi(t)\right>=\ket{P}e^{-Ht}\bra{\phi(0)}$ must
hold and hence, $\ket{P}H=0$ since $\langle P|\phi(0)\rangle=1$, and
thus $H(a_i,b_i,a_i^\dagger\to1,b_i^\dagger\to1)=0$.

Next, we introduce ladder operator coherent states, familiar from
many-particle quantum mechanics. The right eigenstates of the predator
annihilation operator $a_i$ with eigenvalue $\alpha_i$ are
$\bra{\alpha_i}=\exp(-|\alpha_i|^2/2+\alpha_ia_i^\dagger)\bra{0}$,
which can be easily checked by inserting into
$a_i\bra{\alpha_i}=\alpha_i\bra{\alpha_i}$. These states are
overcomplete in the sense that
$\int\prod_id\alpha_i^*d\alpha_i\bra{\alpha_i}\ket{\alpha_i}=\pi$. An
analogous set of right eigenstates can be introduced for the prey
annihilation operator $\beta_i\bra{\beta_i}=\beta_i\bra{\beta_i}$. By
repeatedly inserting the over-completeness relation of both sets of
states into the time-dependent observable~\eqref{eq:observables}, and
following the analysis done in reference \cite{Tauber2012} (described
more generally in reference \cite{Tauber2005}), we arrive at a path
integral expression for calculating averages:
\begin{equation}
  \label{eq:pathintavg}
  \left<O(t)\right>=N^{-1}\int\prod_id\alpha_i^*d\alpha_id\beta_i^*
  d\beta_iO(\cu{\alpha_i},\cu{\beta_i})\exp
  (-S[\cu{\alpha_i^*},\cu{\alpha_i},\cu{\beta_i^*},\cu{\beta_i},t])\;.
\end{equation}
The normalization is determined by calculating the average of the
identity operator
$N=\int\prod_id\alpha_i^*d\alpha_id\beta_i^*d\beta_ie^{-S}$. The 
coherent-state path integral 'action' (the exponential weight in the 
path integral) then becomes
\begin{equation}
  \label{eq:action}
  \begin{split}
    S[\cu{\alpha_i^*},&\cu{\alpha_i},\cu{\beta_i^*},\cu{\beta_i},t]=
    \sum_i\biggl(-\alpha_i(t)-\beta_i(t)-\overline{n}_0\alpha_i^*(0)
    -\overline{m}_0\beta_i^*(0)\\
    &+\int_0^tdt'\biggl[\alpha_i^*\frac{\partial\alpha_i}{\partial t'}
    +\beta_i^*\frac{\partial\beta_i}{\partial t'}+\mu(\alpha_i^*-1)
    \alpha_i+\sigma\Bigl(1-\sum_kf_{ik}\beta_k^*\Bigr)\beta_i^*\beta_i\\
    &+\sum_j\lambda_{ij}\Bigl(\beta_j^*-\sum_kf_{ik}\alpha_k^*\Bigr)
    \alpha_i^*\alpha_i\beta_j\biggr]\biggr)\;.
  \end{split}
\end{equation}
The terms in which the fields explicitly depend on the final time stem 
from the projection state and can be safely ignored for averages and 
correlation functions that do not explicitly depend on these times, as 
is the case here. The variables $\overline{n}_0$ and $\overline{m}_0$ 
respectively represent the average initial number of prey and predator 
particles in each subspecies, and originate in the initial Poisson 
distribution~\eqref{eq:poisson-distrib}.

The classical equations of motion for the fields $\alpha_i^*$,
$\alpha_i$, $\beta_i^*$ and $\beta_i$ are determined by using the
steepest-descent method, i.e. the minimum of $S$ with respect to the
fields. Hence we set the variation of $S$ to zero:
\begin{align}
  \frac{\delta S}{\delta\alpha_i}=0=&-\frac{\partial \alpha_i^*}
  {\partial t}+\mu(\alpha_i^*-1)+\sum_j\lambda_{ij}\Bigl(\beta_j^*-
  \sum_kf_{ik}\alpha_k^*\Bigr)\alpha_i^*\beta_j\;,\label{eq:eom_alpha}\\
  \frac{\delta S}{\delta\beta_i}=0=&-\frac{\partial\beta_i^*}
  {\partial t}+\sigma\Bigl(1-\sum_kf_{ik}\beta_k^*\Bigr)\beta_i^*+
  \sum_j\lambda_{lj}\Bigl(\beta_i^*-\sum_kf_{jk}\alpha_k^*\Bigr)
  \alpha_j^*\alpha_j\;,\label{eq:eom_beta}\\
  \frac{\delta S}{\delta\alpha_i^*}=0=&\frac{\partial\alpha_i}
  {\partial t}+\mu\alpha_i+\sum_j\lambda_{ij}\Bigl(\beta_j^*-\sum_k
  f_{ik}\alpha_k^*\Bigr)\alpha_i\beta_j-\sum_{kj}\lambda_{ij}f_{ki}
  \alpha_k^*\alpha_k\beta_j\;,\label{eq:eom_alphas}\\
  \begin{split}\frac{\delta S}{\delta\beta_i^*}=0=&\frac{\partial
  \beta_i}{\partial t}+\sigma\Bigl(1-\sum_kf_{ik}\beta_k^*\Bigr)
  \beta_i-\sigma\sum_kf_{ki}\beta_k^*\beta_k+\sum_j\lambda_{ji}
  \alpha_j^*\alpha_j\beta_i\\&+\sum_j\lambda_{ji}\Bigl(\beta_i^*-
  \sum_kf_{ik}\alpha_k^*\Bigr)\alpha_i^*\alpha_i\;.\label{eq:eom_betas}
  \end{split}
\end{align}

Equations~\eqref{eq:eom_alpha} and~\eqref{eq:eom_beta} are readily
solved by $\alpha_i^*=1=\beta_i^*$, a consequence of probability
conservation. Equations~\eqref{eq:eom_alphas} and~\eqref{eq:eom_betas}
then yield the classical equations of motion for the fields $\alpha_i$
and $\beta_i$. Since the predator and prey subspecies counts are
$a_i(t)=\ket{P}n_i\bra{\phi(t)}=\alpha_i(t)$ and
$b_i(t)=\ket{P}m_i\bra{\phi(t)}=\beta_i(t)$, we arrive at the coupled
mean-field equations for our system:
\begin{align}
  \frac{\partial a_i(t)}{\partial t}=&-\mu a_i(t)+\sum_{jk}\lambda_{kj}
  f_{ki}a_k(t)b_j(t)\;,\label{eq:eom_fielda}\\
  \frac{\partial b_i(t)}{\partial t}=&\sigma\sum_kf_{ki}b_k(t)-\sum_j
  \lambda_{ji}a_j(t)b_i(t)\;.\label{eq:eom_fieldb}
\end{align}
These equations look very similar to the standard LV rate 
equations~\eqref{eq:lv_mean_field}. In fact, setting 
$f_{ij}=\delta_{ij}$ and $\lambda_{ij}=\lambda\delta_{ij}$ yields the
standard LV mean-field rate equations for each subspecies $i$.

\subsubsection{Steady-State Solutions.}
\label{sec:uniform-distribution}

Steady-state solutions of the mean-field 
equations~\eqref{eq:eom_fielda} and~\eqref{eq:eom_fieldb} are 
determined by setting the time derivatives to zero: 
$\partial a_i(t)/\partial t=0=\partial b_i(t)/\partial t$. Therefore, 
the steady-state particle counts can always be found by numerically 
solving the coupled implicit equations
\begin{align}
  \label{eq:steadystate-eq-a}
  \mu a_i&=\sum_{jk}\lambda_{kj}f_{ki}a_kb_j\;,\\
  \label{eq:steadystate-eq-b}
  \sigma\sum_kf_{ki}b_k&=\sum_j\lambda_{ji}a_jb_i\;,
\end{align}
using a self-consistent, iterative approach.

In the special case of a uniform inheritance distribution $f_{ij}=1/N$, 
the steady-state counts can be found exactly. In this situation, there 
are no correlations between the parent and offspring particle 
efficiencies, and the right-hand side of 
equations~\eqref{eq:steadystate-eq-a} becomes independent of the index
$i$. Consequently, the number of predators in bin $i$ is constant and
independent of $i$, whence $a_i={\rm const}=A$. 
Equation~\eqref{eq:steadystate-eq-b} can be rewritten as
\begin{equation}
  \label{eq:expression-for-bi}
  \frac{b_i}{\sum_kb_k}=\frac{\sigma}{AN}\frac{2N}{\sum_j(i+j+1)}.
\end{equation}
Summing both sides over $i$ and using $\sum_j1=N$ and
$\sum_jj=N(N-1)/2$ gives
\[\frac{AN}{2\sigma}=\sum_i\frac{1}{i+\frac{N+1}{2}}\\;.\]
Using a difference equation involving the digamma function
$\psi(x+N)-\psi(x)=\sum_i\frac{1}{i+x}$ yields
\[\frac{AN}{2\sigma}=\psi\Bigl(\frac{3N+1}{2}\Bigr)-\psi\Bigl(
\frac{N+1}{2}\Bigr)\;.\]
In order to find a useful, approximate value of the constant $A$, we 
rewrite this expression in the form
\[\frac{AN}{2\sigma}=\ln\Bigl(\frac{3N+1}{N+1}\Bigr)+\frac{1}{3N+1}-
\frac{1}{N+1}-\sum_{n=1}^\infty\frac{2^{2n-1}B_{2n}}{n}\Bigl[\frac{1}
{(3N+1)^{2n}}-\frac{1}{(N+1)^{2n}}\Bigr]\,,\]
where we have used the asymptotic series expansion of the digamma 
function
\[\psi(x)=\ln x+\frac{1}{2x}-\sum_{n=1}^\infty\frac{B_{2n}}{2nx^{2n}}\]
($B_k$ is the $k$-th Bernoulli number). Hence, in the limit of large
$N$, the constant simplifies to
\[\lim_{N\to\infty}\frac{AN}{2\sigma}=\ln 3\;.\]
Defining the subspecies densities as $\rho_{a,i}=a_i/\sum_ja_j$ and
$\rho_{b,i}=b_i/\sum_jb_j$, and using
equation~\eqref{eq:expression-for-bi}, as well as the definition of
the efficiency bins $\eta_i=(i+1/2)/N$, we finally arrive at
\begin{equation}
  \label{eq:uniform-sol}
  \rho_{a}=\frac{1}{N}\;,\quad 
  \rho_{b,i}=\frac{2}{N\ln3}\frac{1}{1+2\eta_i}\;,
\end{equation}
which is valid in the limit of large $N$. Hence, the predator density
becomes constant and independent of the subspecies index $i$. The prey
density exhibits a selection bias towards low values of the efficiency
$\eta$.

\subsection{Population Distributions from Simulations}
\label{sec:popul-distr}

We are now ready to perform Monte Carlo simulations of our system. Our
main goal in this section is to extract the predator and prey
population distributions as a function of the particle efficiencies. 
To this end, we introduce efficiency bins $\eta_i=(i+1/2)/N$, 
$i=0,...,N-1$ in complete analogy to the derivation of the master 
equation in section~\ref{sec:master-equation}. We then count the 
number of particles $a_i$ and $b_i$ in the interval
$\bigl[\eta_i-1/(2N),\eta_i+1/(2N)\bigr)$ and calculate the
densities $\rho_{A,i}=a_i/\sum_ja_j$ and $\rho_{B,i}=b_i/\sum_jb_j$. 
The resulting histograms approximate the population distributions as a 
function of the efficacies.

Our simulations start by assigning all particles an initial predation
efficiency of $\eta_{A/B}=0.5$. Hence the population distributions for
$t=0$ exhibit a sharp peak at $\eta=0.5$ and are zero everywhere else.
This choice is mainly due to computational convenience, since the
final steady-state population distributions do not depend on the
initial state of the system. We checked this statement by varying the
initial distribution of particles in efficiency space. There of course
exist initial conditions in which the probability of one or both of
the species to go extinct is rather high. Since we are interested only
in steady states that exhibit species coexistence, we exclude those
initial conditions from our considerations. The reproduction and
mortality rates are both set to $\sigma=\mu=0.5$.

Spatial as well as intrinsic temporal correlations in stochastic 
simulations renormalize the results relative to the mean-field 
predictions given by equations~\eqref{eq:steadystate-eq-a} 
and~\eqref{eq:steadystate-eq-b}. A comparison of our data for
zero-dimensional, non-spatial systems with the results taken in
spatially extended systems allows us to disentangle the effects of 
purely temporal and spatial correlations.

We let the system and thus the population distributions evolve over
time, via random sequential Monte Carlo updates. One Monte Carlo step
is complete when, on average, each particle in the system has been
selected once. The predator and prey populations optimize their
predation efficiency over many generations. In each simulation run, 
we wait until the population distributions have reached their
(quasi-) steady-state shapes. Predators benefit from a higher efficacy
value, because their average interaction and thus reproduction rate is
enhanced. Hence, a predator with a high $\eta$ is more likely to have
more offspring, compared to a low-$\eta$ predator particle, which in
turn inherits this high $\eta$ value. This yields an overall
optimization of the predator population towards high efficacies. Prey 
particles on the other hand benefit from low predation efficiency 
values, because their average lifetime is longer than for individuals 
with high $\eta$. Hence a reduced $\eta$ value yields a larger number
of prey offspring particles and, accordingly, the same optimization 
as for the predator population occurs, only towards low $\eta$ values. 
This dynamic, evolutionary optimization finally leads to a 
steady-state efficacy distribution among the individual particles when
the distance of the population maxima from the efficiency edges 
$\eta=0,1$ is balanced by the finite width of the inheritance 
distribution.
\begin{figure}[tbp]
  \centering
  \includegraphics[width=0.9\columnwidth]{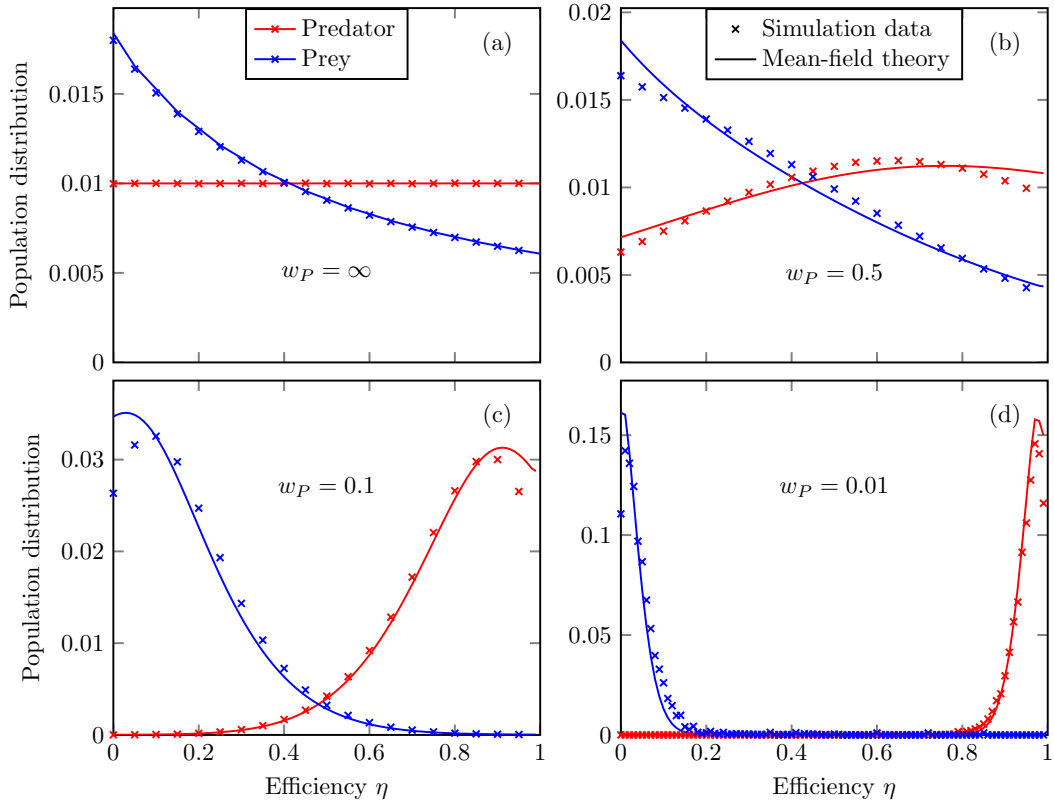}
  \caption[Population distributions for various values of the 
    inheritance distribution width $w_P$.]{Population distributions 
    for various values of the inheritance distribution width $w_P$. 
    The red and blue curves respectively indicate the predator and 
    prey populations as functions of the efficiency. Curves with 
    $\times$ markers stem from zero-dimensional (well-mixed) 
    simulations, while the solid lines show the mean-field predictions.
    (a) Population distribution for a uniform inheritance distribution
    with $w_P=\infty$. The prey population displays an inherent 
    selection bias towards low $\eta$, while the predator population 
    is flat. The mean-field prediction~\eqref{eq:uniform-sol} exactly 
    agrees with the simulation data. (b) Population distribution for a 
    broad inheritance distribution with $w_P=0.5$. The inherent 
    selection bias of the prey population is still visible, but 
    overlaid with the dynamic optimization towards low $\eta$. The 
    predator population optimizes towards higher $\eta$ and shows a 
    maximum around $\eta\approx0.65$. Our numerical mean-field model
    solution is in qualitative agreement. (c) Population distributions
    for a narrow inheritance distribution with $w_P=0.1$. Both 
    predator and prey populations are optimized towards high and low 
    values of $\eta$ with maxima at $\eta\approx0.9$ and 
    $\eta\approx0.1$, respectively. Mean-field theory over-estimates
    the optimization effects and places the population maxima slightly
    closer towards the efficiency extrema. (d) Population distribution
    data for a sharply peaked inheritance distribution with 
    $w_p=0.01$. The maxima move even closer to the edges of the
    efficiency range. The results represent an ensemble average over
    $1000$ realizations.}
  \label{fig:popdistrib}
\end{figure}
\begin{figure}[tbp]
  \centering
  \includegraphics[width=0.9\columnwidth]{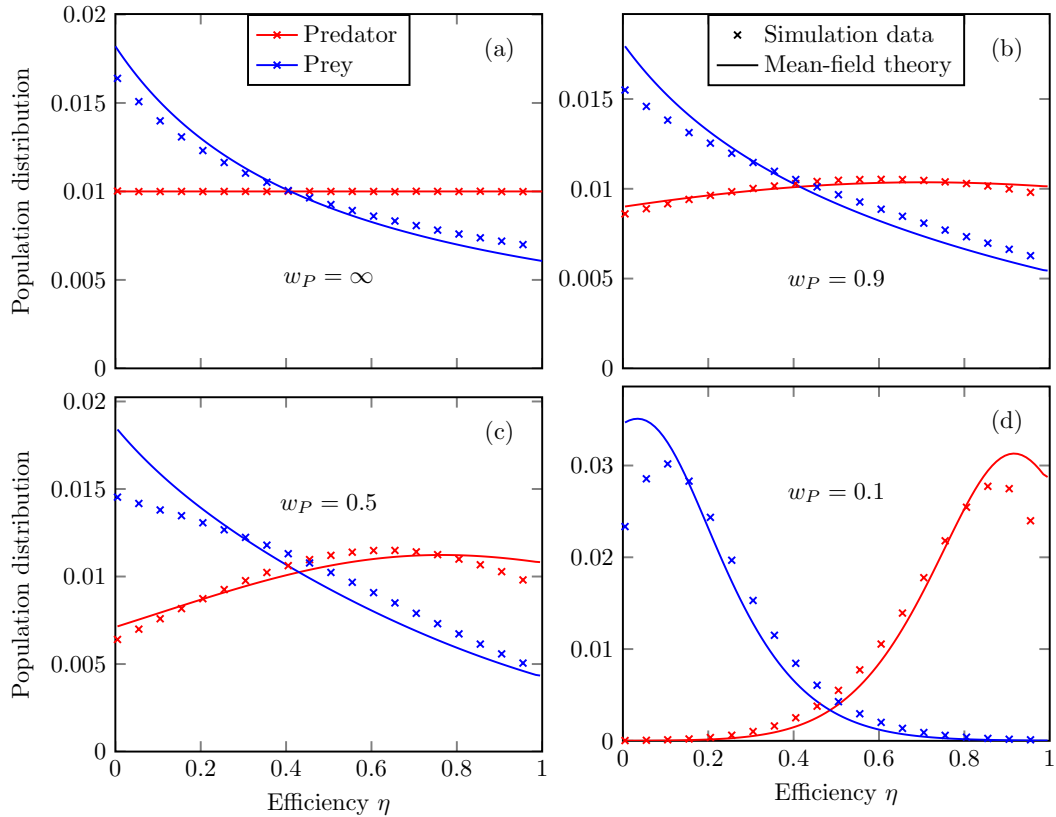}
  \caption[Population distributions for various values of the
    inheritance distribution width $w_P$ from two-dimensional lattice
    simulations.]{Population distributions for various values of the
    inheritance distribution width $w_P$ from two-dimensional lattice
    simulations. The red and blue curves respectively indicate the
    predator and prey populations as a function of the efficiency. 
    Curves with $\times$ markers stem from simulations, while the 
    solid lines show the results from mean-field theory. 
    (a) Population distribution for a uniform inheritance distribution
    with $w_P=\infty$. The mean-field prediction for the prey
    population~\eqref{eq:uniform-sol} ignores spatial correlations and
    thus over-estimates the prey selection bias in this case. (b-d)
    Population distributions for broader inheritance distributions
    with $w_P=0.9,0.5,0.1$, respectively. The mean-field prediction
    deviates more strongly from the simulation data than in the case
    of well-mixed zero-dimensional simulations. The results represent
    an ensemble average over $10000$ realizations.}
  \label{fig:popdistrib_spatial}
\end{figure}

Figure~\ref{fig:popdistrib} displays the population distributions as 
functions of the efficiency $\eta$ for various values of the
inheritance distribution width $w_P$. The special case of an infinite
width $w_P=\infty$ is shown in figure~\ref{fig:popdistrib}(a). In this
situation, no correlation exists between the efficacies of a parent 
and its offspring, and the efficiency assignment during reproduction 
is completely random. Consequently, the predator population 
distribution is flat and independent of $\eta$, as predicted by our 
mean-field theory result~\eqref{eq:uniform-sol}. The prey population 
distribution shows an inherent selection bias towards low values of 
$\eta$. A low efficiency for a prey particle means that it is more 
likely to live longer than another individual with higher efficiency, 
according to our formula for the predation 
rate~\eqref{eq:lambda_efficiencies}. Hence, at any given time when the
system is in the (quasi-) steady state, there needs to be a higher
number of prey particles in the low-efficiency bins than in the higher
ones. This result and the simulation data agree perfectly with our
mean-field theory result~\eqref{eq:uniform-sol} as well, without any
fit parameters.

For non-uniform inheritance distribution, evolutionary optimization of
the predator and prey populations takes place. 
Figure~\ref{fig:popdistrib}(b) shows the population distribution for 
$w_P=0.5$. In this case, the effects of the inherent prey selection 
bias are still apparent and no clear prey population maximum is 
visible. The predator population exhibits a maximum at 
$\eta\approx0.65$ due to the balancing of dynamic optimization and the
finite width of the inheritance distribution. The numerical,
self-consistent solution of our mean-field equations agrees 
qualitatively with the simulation data, but over-estimates the effects 
of optimization. At an even smaller inheritance distribution width of 
$w_P=0.1$, the predator and prey population distributions, displayed 
in figure~\ref{fig:popdistrib}(c), form clear maxima at high and low 
values of $\eta$, respectively. Again, the numerical mean-field 
predictions over-estimate optimization effects and place the 
population maxima nearer to the efficacy edges $\eta=0,1$. A sharply 
peaked inheritance distribution with $w_P=0.01$ yields population 
maxima even closer to the edges of the efficiency range, as shown
in~\ref{fig:popdistrib}(d).

Spatially extended, two-dimensional lattice simulations yield
quantitatively slightly different predator and prey population
distributions. Emerging spatial correlations influence the results as
shown in figure~\ref{fig:popdistrib_spatial}. Since mean-field theory
ignores spatial correlations, our solution already over-estimates the
prey selection bias in the two-dimensional model, but is still
qualitatively correct. A similar trend is noticeable for finite values
of the inheritance distribution width $w_P$. Note that spatial
correlations lead to less sharply peaked population distributions than
in the case of non-spatial simulations. Hence we may conclude that
intrinsic, temporal correlations and spatial correlations both induce
a smoothening of sharp features in the population distributions.

\section{Spatial vs. Demographic Variability}
\label{sec:spat-vs.-demogr}

We now introduce \emph{quenched spatial} randomness in addition to 
demographic variability, which we discussed in the last section. We 
wish to clarify the relative importance of both types of variability 
in the interaction rate on the evolutionary optimization dynamics of
our two-species LV predator-prey system. To this end, we need to 
introduce a new control parameter $\zeta$ that allows us to tune the 
relative influence of environmental and demographic randomness.

We model environmental variability by introducing a new lattice 
site-dependent quenched random variable, the spatial efficiency 
$\eta_S$, similar as in our discussion for purely environmental 
variability in section~\ref{sec:envir-vari}. Before the start of a new
simulation run, the environment is generated by assigning a value to 
this variable on each lattice site, drawn from a Gaussian distribution
of width $w_S$, centered around a value of $\overline{\eta_S}=0.5$ and
truncated to the interval $[0,1]$. The distribution width $w_S$ is a
model parameter and provides a measure of spatial variability similar
to the mutation probability discussed previously. The rate at which an
interaction between two specific predator and prey individuals occurs
on a given lattice site is now a function of $\eta_S$, $\eta_A$, and
$\eta_B$:
\begin{equation}
  \label{eq:spatial-demographic-predation-rate}
  \lambda=\zeta\eta_S+(1-\zeta)\frac{\eta_A+\eta_B}{2}\;.
\end{equation}
The spatial influence parameter $\zeta$ varies between $0$ and $1$ and
smoothly tunes between purely spatial and individual variabilities.

Here, the system consists of a square lattice with $128\times 128$ 
sites and periodic boundary conditions. We did not discern any 
finite-size effects already at this lattice size. Predator and prey 
particles perform unbiased random walks on this lattice, with a 
per-step probability of one. Thus all rates in the system are to be 
understood as measured in units of the diffusivity $D$. The predation,
reproduction, and death reactions occur on-site, without per-site 
particle number restrictions. The predator extinction transition, 
present in systems with site restrictions, is thus absent
here~\cite{Antal2001,Mobilia2006a,Washenberger2007}. The prey
reproduction and predator death rates are fixed and both set to
$\sigma=\mu=0.5$. Predator and prey particles are initially distributed
at random throughout the lattice, with average densities 
$\rho_A=\rho_B=1$. Similar to the zero-dimensional case discussed in 
section~\ref{sec:popul-distr}, the initial individual efficiencies of 
our particles are set to $\eta_A=\eta_B=0.5$, but the final 
steady-state properties turn out not to depend on these initial values.
The simulation proceeds via random sequential Monte Carlo updates, 
where during each iteration a random particle is selected and moved to 
a randomly chosen neighboring site. The particle is then allowed to 
perform a reproduction reaction if it is a prey, or a predation and
subsequent mortality reaction if it is a predator, with the assigned
rates. During a predation reaction, the rate is calculated according
to equation~\eqref{eq:spatial-demographic-predation-rate}. Again, one
Monte Carlo step (MCS) is complete when, on average, each particle in
the system has been selected once.

\subsection{Steady-State Particle Density}
\label{sec:steady-state-part}

\begin{figure}[tbp]
  \centering
  \includegraphics[width=0.9\columnwidth]{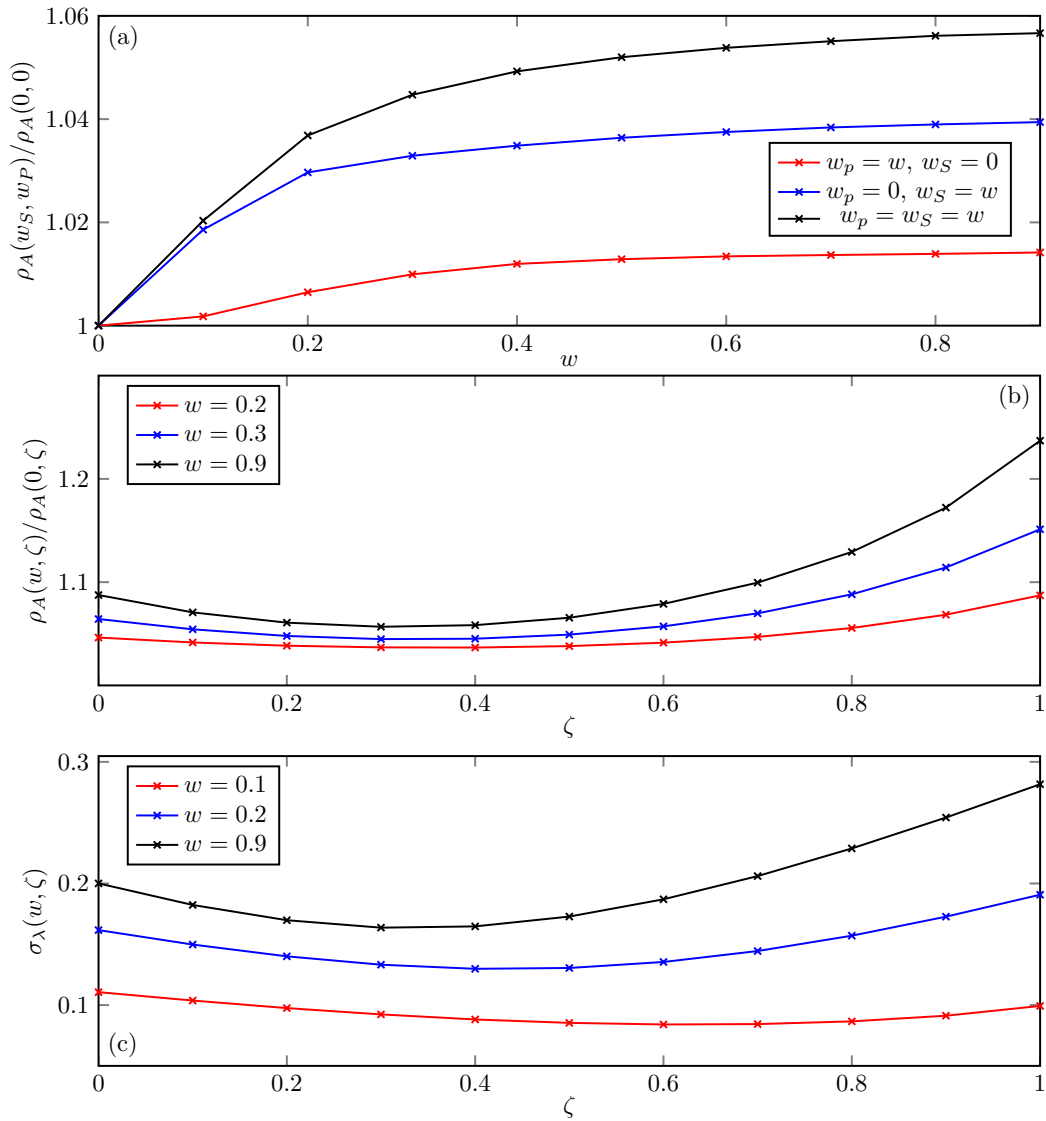}
  \caption{The (quasi-)steady state predator density as a function of
    spatial and individual variability, as well as spatial influence. 
    (a) The normalized (quasi-)steady state predator density $\rho_{A}$
    as a function of variability $w$ for $\zeta=0.3$ for the cases of 
    purely individual ($w_P=w$, $w_S=0$), purely spatial ($w_P=0$, 
    $w_S=w$), and equal variabilities ($w_P=w_S=w$). (b) The normalized
    predator density as a function of $\zeta$ for equal variabilities 
    $w_P=w_S=w$ for $w=0.2$, $w=0.3$, and $w=0.9$. For all values of 
    $w$, a remarkable minimum is observed. (c) The standard deviation 
    of the predation rate $\sigma_\lambda$ for the same cases as in (b), 
    calculated via error propagation from the spatial and individual 
    predation efficiency distributions also shows the minimum.}
  \label{fig:densitysigma}
\end{figure}
We measure the steady-state particle density as a function of the
individual variability $w_P$, the spatial variability $w_S$ and the
spatial importance factor $\zeta$. During each simulation realization
we let the system run for 700 MCS to reach the stationary state and
subsequently average the predator and prey particle densities over an
additional 300 MCS. The resulting data is then averaged over an
ensemble of 10000 realizations per parameter combination. The
investigated ranges for the parameters where $w_P,w_s\in[0,0.9]$ and
$\zeta\in[0,1]$. Figure~\ref{fig:densitysigma}(a) shows the normalized
predator population change as a function of variability for purely
individual ($w_P=w$, $w_S=0$), purely spatial ($w_P=0$, $w_S=w$), and
equal ($w_P=w_S=w$) variabilities, with $\zeta=0.3$ (the location of 
the minimum discussed below). In the case of purely individual
variability we observe a small population increase of $\approx 1.5\%$. 
Purely spatial variability leads to a slightly larger increase of 
$\approx 4\%$, and the mixed case yields the largest increase of just 
below $6\%$. This hierarchy holds true for all values of $\zeta$. 
Figure~\ref{fig:densitysigma}(b) displays the population increase for 
the mixed case, as a function of $\zeta$ for different values of $w$. 
The purely individual ($\zeta=0$) and purely spatial ($\zeta=1$) 
efficiency cases yield local maxima in the population increase, whence 
we observe a remarkable minimum for all values of $w$ for intermediate
values of $\zeta$. Purely spatial efficiency leads to the highest 
observed population density, an increase of just under $25\%$ over the 
non-disordered system. Purely individual rates
yield a moderate increase of $~8\%$. Figure~\ref{fig:densitysigma}(c)
shows the standard deviation for the predation rates $\lambda$,
$\sigma_\lambda=
\sqrt{\zeta^2\sigma_S^2+(1-\zeta)^2(\sigma_A^2+\sigma_B^2)/2}$, as a 
function of $\zeta$ and $w$. Since the spatial and individual predation
efficiency values are truncated to the interval $[0,1]$, the standard 
deviation of their actual distribution is different from the 
variability measure $w$, and therefore needs to be calculated from
simulation data on the population distributions in efficiency space,
and the distribution of spatial efficiency values on the lattice. The
standard deviation follows a similar shape as compared to the
population density shown in figure~\ref{fig:densitysigma}(b); in
particular the two local maxima at $\zeta=0$ and $\zeta=1$, as well as
the minimum in between are reflected here.

Our data clearly demonstrate that the population density increase is 
primarily a monotonic function of the overall variance of the predation 
rate $\lambda$. The two types of variability do not simply contribute 
additively or multiplicatively, since the evolutionary dynamics in the
demographic variability renders the relationship more complex. The
disproportionate increase of the population densities for $\zeta=1$ 
over $\zeta=0$, compared to the standard deviation, also leads us to
conclude that the effect of spatial variability is markedly more 
pronounced as compared to demographic variability.

In section~\ref{sec:uniform-distribution}, we observed that the
evolutionary dynamics inherent to our model of demographic variability
leads to optimization of the population distributions in efficiency
space for \emph{low} values of the variability $w_P$. This becomes
progressively weaker for higher $w_P$. Here, we observe a population
increase for \emph{high} variability, and a very weak to non-existent
increase for lower $w$. Hence, we argue that the optimization of
population distributions in efficiency space is essentially
\emph{neutral} towards the overall species densities (at least in the
context of our model). The net benefit of optimizing the predator
population towards high values of the individual efficiency and the
prey populations towards low efficacies is almost zero. The
optimization is however crucial for the survival of either species
during their competitive co-evolution, reminiscent of an arm's race
scenario.

\subsection{Correlation Lengths and Decay Time}

\begin{figure}[tbp]
  \centering
  \includegraphics[width=0.9\columnwidth]{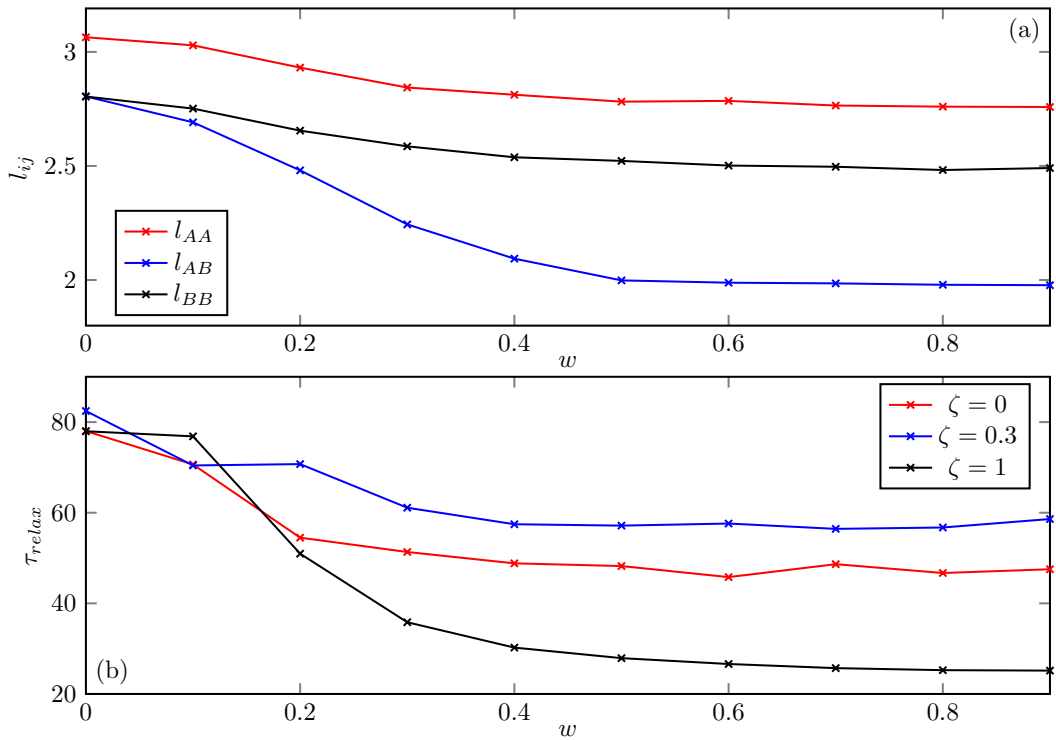}
  \caption{Correlation lengths and relaxation time as a function of
    variability. (a) The autocorrelation lengths $l_{AA}$ and $l_{BB}$
    as well as the cross-correlation length $l_{AB}$ from the species
    autocorrelation functions for $\zeta=0.6$. (b) The predator
    density relaxation time $\tau_{relax}$ toward the quasi-stationary
    state for different values of $\zeta$.}
  \label{fig:corrlengths}
\end{figure}

We calculate correlation lengths by evaluating the spatial density 
correlation functions
\begin{equation}
  \label{eq:corrfunc}
  C_{ij}(x)=\langle\rho_i(0)\rho_j(\vec{x})\rangle
  -\langle\rho_i\rangle\langle\rho_j\rangle,
\end{equation}
where the indices $i$ and $j$ stand for either species $A$ or $B$, 
and the angular brackets indicate an average over all lattice sites 
as well as an ensemble average over $10000$ realizations. The 
single-species auto-correlation functions $C_{AA}$ and $C_{BB}$ 
display a simple exponential decay $C_{ii}(x)\propto\exp(-x/l_{ii})$, 
from which we extract the correlation length via a numerical 
derivative $l_{ii}=-d\ln C_{ii}(x)/dx$. The species cross-correlation 
function $C_{AB}$ is negative for small $x$, has a positive maximum 
at intermediate $x$, and decays to zero for large $x$. We numerically
extract the position of this maximum, the typical distance between
predator and prey particles $l_{AB}$. Figure~\ref{fig:corrlengths}(a)
shows the correlation lengths $l_{AA}$ and $l_{BB}$, as well as the
typical distance $l_{AB}$ as a function of $\zeta$ for $w=0.9$. These
characteristic lengths decrease with increasing variability, which 
indicates that the particles in the system are packed more densely. 
In reference~\cite{Dobramysl2008} we argued that environmental
variability leads to the formation of safe havens for prey, where the
predation rate is very small and prey can proliferate. The predator
particles then feed off the prey particles that diffuse away from the 
activity patches, yielding the observed compression of the system. In
reference~\cite{Dobramysl2013} we observed that a similar mechanism
occurs in the presence of demographic variability, but here these
activity patches are due to highly optimized low-efficiency prey
particles proliferating and thus ephemeral. Consequently the effect of
demographic variability on the steady-state densities is smaller than
the influence of environmental variability.

We additionally investigated the relaxation properties of the LV
system in the presence of both types of variability. To this end, we
Fourier-transformed time traces of the predator density [for an
example see figure~\ref{fig:lv_spatial_stochastic_osc}(a)] and fitted
a Gaussian function to the resulting peak. The peak width is then
inversely proportional to the relaxation time
$\tau_{relax}(w,\zeta)$. Figure~\ref{fig:corrlengths}(b) shows a
consistent decrease in the relaxation time of up to a factor of $0.3$
due to the presence of variability.

\subsection{Extinction Statistics}
  
\begin{figure}[tbp]
  \centering
  \includegraphics[width=0.9\columnwidth]{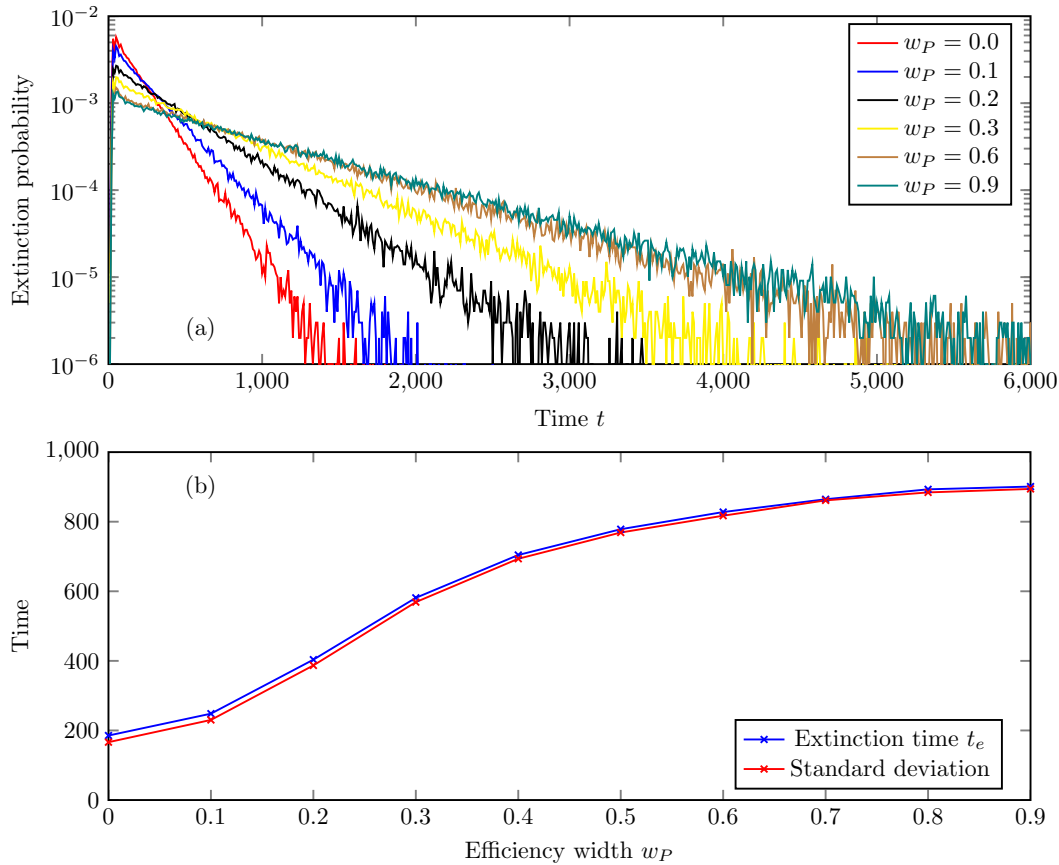}
  \caption[Extinction time probability and mean extinction time in a
    small system of $10 \times 10$ lattice sites as a function of
    individual variability $w_P$.]{Extinction time probability and 
    mean extinction time in a small system of $10 \times 10$ lattice 
    sites as functions of individual variability $w_P$. Only individual
    efficiencies are considered here, $\zeta=0$. (a) Normalized
    extinction event histograms as functions of time for different
    widths of the efficiency inheritance distribution $w_P$. In the
    case of zero variability $w_P=0$, extinction events are mostly
    confined to the time regime $t<1000$. For higher values of $w_P$
    the tail of the extinction event distribution moves to longer
    times and becomes increasingly broader. (b) The mean extinction
    time $t_e$ (blue) shows a more than four-fold increase as a
    function of the variability $w_P$. Its standard deviation (red)
    has approximately the same value as $t_e$, which is consistent
    with an exponential extinction distribution in the long-time
    limit.}
  \label{fig:extinction}
\end{figure}
In finite stochastic systems with an absorbing state (here, predator
extinction), fluctuations will eventually drive the system into the
absorbing state, as discussed in section~\ref{sec:stoch-simul} and
references~\cite{McKane2005,Mobilia2006a}. This can be exploited to 
gain information about the stability of our model against the 
extinction of either species as a function of the model parameters. 
To this end, we simulated small systems, with a lattice size of 
$10\times10$ sites, until the number of particles of either the prey 
or the predator species reaches zero, and collected the simulation 
time up to this event into extinction time histograms. The normalized 
extinction event count then corresponds to the extinction event 
probability $P_e(t)$.

Figure~\ref{fig:extinction}(a) depicts the extinction data for
selected values of the variability $w_P$. The histograms show that the
extinction probabilities are consistent with an exponential
distribution in the long-time
limit~\cite{Parker2009,Dobrinevski2012}. For increasing variability,
the extinction event distributions become increasingly
broader. Figure~\ref{fig:extinction}(b) shows the mean extinction time
$t_e=\sum_{t=0}^{\infty}tP_e(t)$ and its standard deviation
$\sigma_e=\sqrt{\sum_{t=0}^{\infty}(t-t_e)^2P_e(t)}$ as a function of
the inheritance distribution width $w_P$. The mean extinction time is
enhanced by a factor of up to $\approx 4.5$ due to individual
variability. This, together with the increase in $\sigma_e$, indicates
that a higher number of realizations of our small system survive for
longer times. Hence, we conclude that individual variability renders
our model more robust against extinction.

\section{Conclusions}
\label{sec:conclusions}

In this paper, we have studied and discussed a particular variant of 
the LV model in which we introduced two different kinds of variability 
into the predator-prey interaction rate. In an earlier study, we
investigated the effects of purely spatial variability of the 
predation rate and found a marked increase in the steady-state
population densities of both species; see section~\ref{sec:envir-vari}
and reference~\cite{Dobramysl2008}.

Here, we introduced demographic variability together with evolutionary
dynamics, in which during a reproduction step, the offspring particle
inherits an efficiency close to its parent's value. The resulting
steady-state optimization, discussed in section~\ref{sec:popul-distr},
yields predator and prey populations that are located at high and low
values of the efficiency, respectively. We were able to find good
agreement of the simulation data with our effective subspecies 
mean-field model derived in section~\ref{sec:mean-field-equations}. Our
results show that this population level optimization has negligible 
effects on the overall population densities, but is necessary for 
species survival.

In section~\ref{sec:spat-vs.-demogr}, we discussed our results for a
spatially extended system in which both types of variability,
environmental and demographic, are present. We found that demographic
variability leads to an increase of the steady-state densities of both
species, similar to our previous results for purely spatial randomness
but smaller in magnitude. By investigating correlation functions, we
demonstrated that the system becomes denser, supporting our argument 
that variability causes more localized activity patches, where prey 
proliferate and predators feed off prey that diffuse away from these
patches. Additionally, extinction event histograms show that enhanced
variability renders the system more stable against the extinction of
either species.

This extensive numerical Monte Carlo simulation study of environmental
and demographic variability highlights the importance of randomness on
the dynamics of ecological models. While a simple two-species
predator-prey system has limited predictive power for real ecological
neighborhoods, these results still emphasize the need to investigate
variability in more complex models, such as food webs.

We gratefully acknowledge inspiring discussions with G. Daquila,
E. Frey, J. Phillips, T. Platini, M. Pleimling, B. Schmittmann, and
R.K.P. Zia.

\section*{References}

\end{document}